\documentclass[twocolumn,english,aps,prd,nofootinbib,hidelinks]{revtex4-2}
\usepackage{lmodern}
\usepackage[LGR,T1]{fontenc}
\usepackage[latin9]{inputenc}
\usepackage{xcolor}
\usepackage{babel}
\usepackage{amsmath}
\usepackage{amssymb}
\usepackage[unicode=true]
 {hyperref}

\makeatletter

\DeclareRobustCommand{\greektext}{%
  \fontencoding{LGR}\selectfont\def\encodingdefault{LGR}}
\DeclareRobustCommand{\textgreek}[1]{\leavevmode{\greektext #1}}
\ProvideTextCommand{\~}{LGR}[1]{\char126#1}

\usepackage{enumerate}

\makeatother

\begin{document}
\title{Dilaton-induced variations in Planck constant and speed of light:\vskip1pt
An alternative to Dark Energy}
\author{Hoang Ky Nguyen$\,$}
\email[\ ]{hoang.nguyen@ubbcluj.ro}

\affiliation{{\vskip3pt}Department of Physics, Babe\c{s}-Bolyai University, Cluj-Napoca
400084, Romania\vskip1ptInstitute for Interdisciplinary Research in Science and Education,~{\linebreak}ICISE, Quy Nhon 55121, Vietnam\vskip-9pt}
\date{February 26, 2025}
\begin{abstract}
\vskip2ptWe reveal a novel aspect of scale-invariant actions that
allow matter to couple with a dilaton field:\linebreak The dynamics
of the dilaton can induce variations in the Planck constant $\hbar$
and speed of light $c$.\linebreak Our mechanism for generating variable
$\hbar$ and $c$ in \emph{curved} spacetimes via the dilaton offers
a viable\linebreak alternative account for late-time cosmic acceleration,
bypassing the need for dark energy.
\end{abstract}
\maketitle
\textbf{\emph{Motivation}}---Let us revisit the quantum electrodynamics
(QED) of a spinor field $\psi$ with \emph{inertial} mass $m$ and
charge $e$, coupled with a $U(1)$ gauge vector field $A_{\mu}$
and embedded in a 4-dimensional spacetime described by the Einstein--Hilbert
(EH) action of General Relativity (GR). With $\hbar$ and $c$ made
explicit, the full action reads
\begin{align}
\mathcal{S}_{1} & =\int d^{4}x\sqrt{-g}\,\Bigl[\mathcal{L}_{\text{EH}}+\mathcal{L}_{\text{QED}}\Bigr]\label{eq:S-1}\\
\mathcal{L}_{\text{EH}} & =\frac{c^{3}}{16\pi\hbar G}\,\mathcal{R}\label{eq:L-EH}\\
\mathcal{L}_{\text{QED}} & =i\,\bar{\psi}\gamma^{\mu}\nabla_{\mu}\psi+\frac{e}{\sqrt{\hbar c}}\bar{\psi}\gamma^{\mu}A_{\mu}\psi+m\frac{c}{\hbar}\,\bar{\psi}\psi-\frac{1}{4}F_{\mu\nu}F^{\mu\nu}\label{eq:L-QED}\\
F_{\mu\nu} & :=\partial_{\mu}A_{\nu}-\partial_{\nu}A_{\mu}
\end{align}
The restoration of $\hbar$ and $c$ in these expressions will be
instrumental for our later discussion when considering modified theories
of gravity and matter. For instance, in place of $\mathcal{L}_{\text{EH}}$,
Brans and Dicke proposed an extension of GR known as the Brans--Dicke
(BD) gravity action \citep{BransDicke-1961}:
\begin{equation}
\mathcal{L}_{\text{BD}}=\phi\,\mathcal{R}-\frac{\omega}{\phi}g^{\mu\nu}\partial_{\mu}\phi\partial_{\nu}\phi\label{eq:L-BD}
\end{equation}
Traditionally, the dynamical BD scalar field $\phi$ is regarded as
an inverse (\emph{variable}) Newton gravitational constant $G$. However,
by comparing the right-hand-side of Eq. \eqref{eq:L-EH} with the
first term in the right-hand-side of Eq. \eqref{eq:L-BD}, it becomes
evident that the \emph{combination} $c^{3}/(\hbar G)$, and not merely
$1/G$, should be associated with $\phi$. If the matter Lagrangian
only allows matter to couple minimally with gravity---such as $\mathcal{L}_{\text{QED}}$
in Eq. \eqref{eq:L-QED}---the inertial mass $m$ and charge $e$,
along with $\hbar$ and $c$, are unrelated to $\phi$; in this situation,
$1/G$ can be identified with $\phi$, as initially proposed by Brans
and Dicke.\vskip4pt

However, this relationship between $G$ and $\phi$ could cease to
hold if matter couples \emph{non-minimally} with gravity via the BD
field. In this scenario, the involvement of $\phi$ \emph{in the matter
Lagrangian}, in principle, can cause both $c$ and $\hbar$ to vary
alongside $\phi$ in spacetime, representing a departure from the
standard BD paradigm. The current Letter explores this potentially
far-reaching scenario.\vskip8pt

\textbf{\emph{Derivation}}---For convenience, let us replace the
BD field $\phi$ with a scalar field $\chi$ via the substitution
$\phi:=\chi^{2}$, transforming the BD action into
\begin{equation}
\mathcal{L}_{\chi}=\chi^{2}\,\mathcal{R}-4\omega\,g^{\mu\nu}\partial_{\mu}\chi\partial_{\nu}\chi\label{eq:L-chi}
\end{equation}
Furthermore, we allow the quadratic term $\bar{\psi}\psi$ to couple
\emph{non-minimally} with gravity via the field $\chi$ in the form
$\chi\,\bar{\psi}\psi$. The modified QED Lagrangian is given by
\begin{equation}
\mathcal{L}_{\text{QED}}^{(\chi)}=i\,\bar{\psi}\gamma^{\mu}\nabla_{\mu}\psi+\sqrt{\alpha}\,\bar{\psi}\gamma^{\mu}A_{\mu}\psi+\mu\,\chi\,\bar{\psi}\psi-\frac{1}{4}F_{\mu\nu}F^{\mu\nu}\label{eq:L-chi-QED}
\end{equation}
where the Dirac gamma matrices satisfy $\gamma^{\mu}\gamma^{\nu}+\gamma^{\nu}\gamma^{\mu}=2\,g^{\mu\nu}$,
and the spacetime covariant derivative $\nabla_{\mu}$ acts on the
spinor via vierbein and spin connection. The full action now reads
\begin{align}
\mathcal{S}_{2} & =\int d^{4}x\sqrt{-g}\,\Bigl[\mathcal{L}_{\chi}+\mathcal{L}_{\text{QED}}^{(\chi)}\Bigr]\label{eq:S-2}
\end{align}
All parameters $\omega$, $\alpha$, and $\mu$ in $\mathcal{L}_{\chi}$
and $\mathcal{L}_{\text{QED}}^{(\chi)}$ are dimensionless. The modified
action $\mathcal{S}_{2}$ is scale invariant; i.e., it remains unchanged
under a global Weyl rescaling:
\begin{equation}
g_{\mu\nu}\rightarrow\Omega^{2}g_{\mu\nu};\ \psi\rightarrow\Omega^{-3/2}\psi;\ A_{\mu}\rightarrow A_{\mu};\ \chi\rightarrow\Omega^{-1}\chi
\end{equation}

In the literature, when scale symmetry, also known as dilatation symmetry,
is broken---specifically, when $\chi$ spontaneously acquires a non-vanishing
vacuum expectation value---the process gives rise to a Goldstone
mode conventionally referred to as a dilaton. We will adopt the terminology
of ``dilaton'' for $\chi$ in this Letter. We should note, however,
that the field $\chi$ in our action $\mathcal{S}_{2}$ can vary in
spacetime, hence forfeiting its translational invariance. \vskip4pt

Interestingly, a scale-invariant action of gravity \emph{and} matter,
such as $\mathcal{S}_{2}$, is able to evade observational bounds
on the fifth force, a result established in \citep{Blas-2011,Ferreira-2017}.
One can also add a ``potential'' term $V(\chi)$ to $\mathcal{L}_{\chi}$
provided that the added term respects scale symmetry. For example,
$V(\chi)$ may contain terms like $\chi^{4}$ and $\chi^{-4}(\nabla\chi)^{4}$.
Nevertheless, the inclusion of a scale-invariant potential will not
affect the consideration presented in the rest of this Letter.\vskip8pt

\textbf{\emph{Identification}}---Consider the dilaton $\chi$ as
a slowly varying background field in spacetime. A comparison of $\mathcal{L}_{\chi}$
versus $\mathcal{L}_{\text{EH}}$ and $\mathcal{L}_{\text{QED}}^{(\chi)}$
versus $\mathcal{L}_{\text{QED}}$ results in the following identification:\vspace{-0.25cm}

\begin{equation}
\frac{c^{3}}{16\pi\hbar G}:=\chi^{2};\ \ \ \frac{e}{\sqrt{\hbar c}}:=\sqrt{\alpha};\ \ \ m\frac{c}{\hbar}:=\mu\,\chi\label{eq:identification}
\end{equation}
We will present two alternative routes to meet these identities.\vskip8pt

\textbf{\emph{The Fujii--Wetterich scheme}}: Under the canonical
assumption that $\hbar$ and $c$ are constants, one then obtains
from \eqref{eq:identification}:
\begin{equation}
e=\bigl(\alpha\hbar c\bigr)^{1/2};\ \ \ m_{\chi}=\frac{\mu\hbar}{c}\,\chi\label{eq:e-m-scheme1}
\end{equation}
and\vspace{-.5cm}
\begin{equation}
G_{\chi}=\frac{c^{3}}{16\pi\hbar}\,\chi^{-2}\,.\label{eq:G-scheme1}
\end{equation}
Here, the subscript $\chi$ in $m_{\chi}$ and $G_{\chi}$ underlines
the dependence of $m$ and $G$ on $\chi$. Notably, the gauge charge
$e$ remains \emph{unrelated} to $\chi$. To the best of our knowledge,
this scheme, which results in variable $G$ and \emph{variable mass},
was first laid out by Fujii \citep{Fujii-1982} and Wetterich \citep{Wetterich-1988a,Wetterich-1988b,Wetterich-2013a,Wetterich-2013b,Wetterich-2014}
although they approached it using matter--gravity actions different
from $\mathcal{S}_{2}$. The Planck mass, defined as $m_{\chi}^{\text{Planck}}\,:=$$\sqrt{\frac{\hbar c}{G_{\chi}}}=$$\,\sqrt{16\pi}\frac{\hbar}{c}\,\chi$,
is variable in this Fujii--Wetterich (FW) scheme.\vskip4pt

It is important to note that in this scheme, owing to the relationship
involving $m_{\chi}$ in Eq. \eqref{eq:e-m-scheme1}, the dilaton
can only affect massive particles, viz. the quanta of the spinor field
$\psi$. The dilaton does not impact massless particles, viz. the
quanta of the $U(1)$ gauge vector field $A_{\mu}$. \linebreak

\textbf{\emph{Our scheme}}: There is no \emph{a priori} theoretical
reason to prevent Planck's quantum of action and the speed of light
from associating with the dilaton, however. Instead, it is permissible
to relate $\hbar$ and $c$ to $\chi$ via the following assignments
(with the subscript $\chi$ attached):
\begin{equation}
\hbar_{\chi}:=\hat{\hbar}\ \Bigl(\frac{\chi}{\hat{\chi}}\Bigr)^{-1/2};\ \ \ c_{\chi}:=\hat{c}\ \Bigl(\frac{\chi}{\hat{\chi}}\Bigr)^{1/2}\label{eq:scheme2-assign}
\end{equation}
where $\hat{\hbar}$ and $\hat{c}$ represent the values of $\hbar_{\chi}$
and $c_{\chi}$ at a reference point where $\chi=\hat{\chi}$ (with
$\hat{\chi}\neq0$). Employing Eq. \eqref{eq:identification}, our
assignments unambiguously lead to
\begin{equation}
e=\bigl(\alpha\hat{\hbar}\hat{c}\bigr)^{1/2};\ \ \ m=\frac{\mu\,\hat{\chi}\hat{\hbar}}{\hat{c}}\label{eq:e-m-scheme2}
\end{equation}
and\vspace{-.5cm}
\begin{equation}
G=\frac{\hat{c}^{3}}{16\pi\hat{\chi}^{2}\hat{\hbar}}\label{eq:G-scheme2}
\end{equation}

Importantly, per Eq.$\,$\eqref{eq:e-m-scheme2}, the charge $e$
and inertial mass $m$ are independent of $\chi$. This result is
desirable, as both charge and \emph{inertial} mass---being \emph{intrinsic}
properties of a particle---should be oblivious to external factors,
in particular the \emph{background} dilaton field $\chi$. Befittingly,
in our scheme, rather than making mass variable, \emph{the effect
of the dilaton is translated into $\hbar_{\chi}$ and $c_{\chi}$,
which respectively regulate the quantization and propagation of fields
against a background spacetime.}\vskip4pt

The role of $\chi$ in determining $\hbar$ and $c$ underscores a
stark distinction between the two schemes. As mentioned earlier, in
the FW scheme, the dilaton---by design---can only affect massive
particles while leaving massless particles unaffected. In contrast,
our scheme allows the dilaton---through its involvement with $\hbar$
and $c$---to govern the quantization and propagation of \emph{all
fields}, irrespective of their mass parameters. \emph{The ability
to make the dilaton universally affect all types of fields is a major
virtue of our scheme in comparison to the FW scheme.}\vskip4pt

Moreover, our scheme brings about an additional benefit: not only
is the $U(1)$ gauge charge $e$---which specifies the strength of
electromagnetic interaction---independent of $\chi$, but the Newton
gravitational constant $G$---characterizing the strength of gravitational
interaction---is also independent of $\chi$, as evident in Eq. \eqref{eq:G-scheme2}.
Note that in our scheme, the Planck mass, given by $m_{\chi}^{\text{Planck}}\,:=$\linebreak$\sqrt{\frac{\hbar_{\chi}c_{\chi}}{G}}=$$\,\sqrt{16\pi}\,\frac{\hat{\chi}\hat{\hbar}}{\hat{c}}$,
is constant rather than variable. \footnote{Our Letter does not concern with the hierarchy issue, namely why $m_{\chi}^{\text{Planck}}\gg m$,
or equivalently why $\mu\ll\sqrt{16\pi}$.}\vskip4pt

Two comments are necessary. Firstly, the dimensionless parameters
$\alpha$ and $\mu$ are ``running coupling constants'' in the renormalization
group flow of $\mathcal{L}_{\text{QED}}^{(\chi)}$ when loop corrections
involving $\psi$ and $A_{\mu}$ are included. Thus, although $e$
and $m$ are independent of $\chi$, they can ``run'' as functions
of the momentum scale at which they are measured.\vskip4pt

Secondly, per Eq. \eqref{eq:identification}, the fine structure constant
$\alpha$ is unrelated to $\chi$ and thus does not vary in spacetime.
This criterion is compatible with the theoretical bound that time-variations
in $\alpha$, if caused by a cosmic scalar field, would be of order
$\left|\frac{\delta\,\alpha}{\alpha}\right|<10^{-37}$ since the earliest
stages of galaxy formation \citep{Banks-2002}. (This bound hence
casts doubt on the validity of the observation reported in \citep{Webb-2001}.)\vskip8pt

\textbf{\emph{A mechanism to generate variable $\hbar$ and $c$}}---Consider
an open set surrounding a given point $x^{*}$ on the spacetime manifold.
With the dilaton $\chi$ varying sufficiently slowly, it can be treated
as having a \emph{constant} value within the open set. Utilizing Eq.
\eqref{eq:scheme2-assign} from our scheme, the Planck `constant'
and the speed of light acquire their respective values $h_{\chi}$
and $c_{\chi}$, to be \emph{applicable solely for this region}. In
a tangent frame to the manifold at $x^{*}$, our scheme maps $\mathcal{L}_{\text{QED}}^{(\chi)}$
to an \emph{effective} Lagrangian given by
\begin{align}
\mathcal{L}_{\text{QED}}^{\text{ eff}} & =i\,\bar{\psi}\gamma^{\mu}\nabla_{\mu}\psi+\frac{e}{\sqrt{\hbar_{\chi}c_{\chi}}}\bar{\psi}\gamma^{\mu}A_{\mu}\psi+m\frac{c_{\chi}}{\hbar_{\chi}}\,\bar{\psi}\psi\nonumber \\
 & -\frac{1}{4}F_{\mu\nu}F^{\mu\nu}\label{eq:L-eff}
\end{align}

Within the open set, $\mathcal{L}_{\text{QED}}^{\text{ eff}}$ describes
the quantum electrodynamics of a spinor field with charge $e$ and
inertial mass $m$, specified in Eq. \eqref{eq:e-m-scheme2}, coupled
with a $U(1)$ gauge vector field. It is important to emphasize that,
unlike the original $\mathcal{L}_{\text{QED}}$ in Eq. \eqref{eq:L-QED},
the quantization and propagation of fields in $\mathcal{L}_{\text{QED}}^{\text{ eff}}$
are governed by the effective quantum of action $\hbar_{\chi}$ and
speed of light $c_{\chi}$, which in turn are determined by the dilaton
$\chi$ via Eq. \eqref{eq:scheme2-assign}. More generally, \emph{as
$\chi$ varies across the manifold, different locations then correspond
to separate replicas of the effective Lagrangian }$\mathcal{L}_{\text{QED}}^{\text{ eff}}$\emph{,
each operating with its respective values of $\hbar_{\chi}$ and $c_{\chi}$}.
Conceptually, the dynamics of $\chi$ thus induces variations in $\hbar$
and $c$ across spacetime. \footnote{The co-variability of $\hbar$ and $c$ has been considered in Refs.
\citep{Gupta-2021,Buchalter-2004,Cuzinatto-2022,Gupta-2020a}, \linebreak
although a concrete mechanism for generating their variations in spacetime
is still lacking in these works.}\vskip4pt

We emphasize that although the action described in Eqs. \eqref{eq:L-chi},
\eqref{eq:L-chi-QED} and \eqref{eq:S-2} has been previously explored
in both the contexts of gravitation and particle physics (e.g., in
Refs. \citep{Ferreira-2017,Fujii-1982,Wetterich-1988a,Wetterich-1988b,Wetterich-2013a,Wetterich-2013b,Wetterich-2014,Blas-2011,Ghilencea-2019,Kannike-2017,Karananas-2016,Nishino-2011,Rubio-2014,Rubio-2017,Salvio-2014,Shaposhinikov-2011,Einhorn-2017,Bardeen-1995}),
its capacity to produce variations in $\hbar$ and $c$ in spacetime---as
presented in this Letter thus far---has not been recognized or documented.\vskip8pt

\textbf{\emph{The operational meaning of variable $c$ and $\hbar$}}---In
1911 Einstein originated the idea of a variable speed of light (VSL)
during his formulation of GR \citep{Einstein-1911,Einstein-1912-a,Einstein-1912-b}.
Although he did not pursue this idea further, the possibility of VSL
was revived by Dicke in 1957 \citep{Dicke-1957}, as well as by Moffat,
Albrecht, and Magueijo in the 1990s \citep{Moffat-1992,Magueijo-1999}.
Importantly, as Einstein recognized in \citep{Einstein-1912-a,Einstein-1912-b},
the VSL concept does not contradict the principle of the constancy
of $c$ with respect to local Lorentz boosts; $\ $this is because
the Lorentz symmetry is only required to be valid \emph{locally},
while the value of $c$ may vary from one point to another in spacetime.
Furthermore, it is \emph{incorrect} to dismiss the distinction between
keeping $c$ constant or allowing it to vary as merely a trivial redefinition
of units (i.e., rods and clocks): a varying $c$ leads to a refraction
effect---a physically measurable phenomenon---that impacts the propagation
of light rays, whereas a constant $c$ does not.\vskip4pt

The action $\mathcal{S}_{2}$ is generally covariant, whereas the
action associated with $\mathcal{L}_{\text{QED}}^{\text{ eff}}$ is
locally Lorentz invariant. Our approach is therefore compatible with
the varying--$c$ framework that respects both general covariance
and local Lorentz invariance, as laid out by Magueijo in \citep{Magueijo-2000}.\vskip4pt

A variable $\hbar$---as a function of $\chi$---would affect the
quantization of fields, specifically the commutation relation of position
and momentum for particles, given by
\[
\hat{x}\,\hat{p}-\hat{p}\,\hat{x}=i\,\hbar_{\chi}
\]
In relation to this, a variable $\hbar$ would influence the time
evolution of a quantum state $\bigl|\Psi\bigr\rangle$ according to
\begin{equation}
i\hbar_{\chi}\,\frac{\partial}{\partial t}\bigl|\Psi\bigr\rangle=\hat{H}\bigl|\Psi\bigr\rangle\label{eq:evolution}
\end{equation}
A physical consequence is that, for two locations with different dilaton
values $\chi_{1}$ and $\chi_{2}$, the evolution of a quantum state
occurs at different rates determined by each respective $h_{\chi}$
value. The discrepancy in the clock rates at the two locations gives
rise to a \emph{new} time dilation effect, resulting from the variation
in $\hbar$ induced by the dynamics of $\chi$. We will briefly discuss
this concrete definitive prediction in a later section of the Letter.
\vskip8pt

\textbf{\emph{Illustration using the Hydrogen atom---}}By applying
the variational principle to the spinor field and the gauge vector
field in the effective $\mathcal{L}_{\text{QED}}^{\text{ eff}}$ given
in Eq. \eqref{eq:L-eff}, it is straightforward to derive the Dirac
equation
\begin{align}
\Biggl(i\gamma^{\mu}\partial_{\mu}+\frac{e}{\sqrt{\hbar_{\chi}c_{\chi}}}\gamma^{\mu}A_{\mu}+m\,\frac{c_{\chi}}{\hbar_{\chi}}\Biggr)\,\psi & =0\label{eq:my-Dirac}
\end{align}
and the Maxwell equation
\begin{equation}
\partial_{\nu}F^{\nu\mu}=\frac{e}{\sqrt{\hbar_{\chi}c_{\chi}}}\,\,\bar{\psi}\gamma^{\mu}\psi\,.\label{eq:my-Maxwell}
\end{equation}
For convenience, we will refer to $\psi$ as an electron field and
$A_{\mu}$ as an electromagnetic (EM) field. For a Coulomb potential,
where $A_{0}=-\frac{1}{\sqrt{\hbar_{\chi}c_{\chi}}}\frac{e}{r}$,
$A_{1}=A_{2}=A_{3}=0$, the Dirac equation becomes
\begin{equation}
\Biggl(i\gamma^{\mu}\partial_{\mu}-\gamma^{0}\frac{1}{\hbar_{\chi}c_{\chi}}\frac{e^{2}}{r}+m\,\frac{c_{\chi}}{\hbar_{\chi}}\Biggr)\,\psi=0
\end{equation}
which describes the motion of an electron in a hydrogen atom. The
Bohr radius of its groundstate is given by
\begin{align}
a_{B} & =\frac{\hbar_{\chi}}{\alpha\,mc_{\chi}}=\frac{1}{\alpha\,\mu\,\chi}\propto\chi^{-1}\label{eq:Bohr-radius-ours}
\end{align}
which is reciprocal to $\chi$. More generally, it can be anticipated---from
the ground of dimensionality---that the lengthscale $l$ of any physical
process occurring within the open set of a constant $\chi$ is also
reciprocal to $\chi$, viz.
\begin{equation}
l\propto\chi^{-1}
\end{equation}
\vskip4pt

The energy level of an (relativistic) electron in a quantum state
$\left|n,\,j\right\rangle $ of a Hydrogen atom is a well-established
result; it is given by \citep{Greiner-RQM-book} \footnote{Beyond the Dirac theory, the two states $^{2}S_{1/2}$ and $^{2}P_{1/2}$
of the Hydrogen atom become non-degenerate due to radiative corrections,
a physical effect known as the Lamb shift. Their energy difference
is given by $\frac{1}{6\pi}\alpha^{5}mc^{2}\ln\frac{1}{\pi\alpha}$
\citep{Welton-1948} which is also proportional to $\chi$ due to
$c\propto\chi^{1/2}$. We thank the Referee for suggesting this connection.}
\begin{align}
E_{n}^{j} & =\mathcal{N}_{n}^{j}\,mc_{\chi}^{2}=\mathcal{N}_{n}^{j}\,\mu\hat{\hbar}\hat{c}\,\chi\propto\chi\label{eq:E-nj-ours}
\end{align}
which is proportional to $\chi$. In the above expression, the factor
$\mathcal{N}_{n}^{j}\!$ is equal to ${\scriptstyle {\scriptstyle \left(1+\alpha^{2}\left(n-j-\frac{1}{2}+\sqrt{\left(j+\frac{1}{2}\right)^{2}-\alpha^{2}}\right)^{-2}\right)^{-1/2}}}$
with $n=1,2,3,...$ and $j=\frac{1}{2},\frac{3}{2},\frac{5}{2},...$.\vskip8pt

\textbf{\emph{Anisotropic scaling in the clock rate}}---Induced by
electric dipole, a hydrogen atom in the excited state $\left|n=2,\,j=3/2\right\rangle $
can spontaneously transition to the groundstate $\left|n=1,\,j=1/2\right\rangle $,
as allowed by the selection rule $\Delta j=\pm1$. The energy of the
photon emitted is $\Delta E=E_{n=2}^{j=3/2}-E_{n=1}^{j=1/2}$, and
the frequency of the emitted photon is\vspace{-.1cm}
\begin{align}
\nu & =\frac{\Delta E}{\hbar_{\chi}}=\Bigl(\mathcal{N}_{n=2}^{j=3/2}-\mathcal{N}_{n=1}^{j=1/2}\Bigr)\,\frac{\mu\hat{c}}{\hat{\chi}^{1/2}}\,\chi^{3/2}\propto\chi^{3/2}\label{eq:nu}
\end{align}
Therefore, the propagation of the photon within the open set of constant
$\chi$ has a timescale that behaves as
\begin{equation}
\tau:=\frac{1}{\nu}\propto\chi^{-3/2}\label{eq:tau-photon}
\end{equation}

Similarly, the rate of transition (i.e. Einstein's coefficient) from
the initial state $\left|i\right\rangle =\left|n=2,\,j=3/2\right\rangle $
to the final state $\left|f\right\rangle =\left|n=1,\,j=1/2\right\rangle $
can be found to be \citep{QM-book}:\vspace{-.1cm}

\begin{equation}
A=\frac{4\alpha}{3}\frac{m^{3}c_{\chi}^{4}}{h_{\chi}^{3}}\left\langle \vec{r}_{if}\right\rangle ^{2}\label{eq:decay-1}
\end{equation}
where the matrix element of the electric dipole is $\left\langle \vec{r}_{if}\right\rangle :=\left\langle f|\vec{r}|i\right\rangle $.
Given that $\left\langle \vec{r}_{if}\right\rangle \propto\chi^{-1}$,
$c_{\chi}\propto\chi^{1/2}$, and $h_{\chi}\propto\chi^{-1/2}$, the
transition rate behaves as $A\propto\chi^{3/2}$. Thus, the half-life
of the decay process for an unstable quantum system scales as\vspace{-.1cm}
\begin{equation}
\tau:=\frac{1}{A}\propto\chi^{-3/2}
\end{equation}
in alignment with Eq. \eqref{eq:tau-photon}. Although this result
was derived through illustrations involving a hydrogen atom, the time
scaling behavior is generic owing to the time evolution of quantum
states, given in Eq. \eqref{eq:evolution}. With the Hamiltonian scales
similarly to energy (i.e., $\hat{H}\propto\chi$ per Eq. \eqref{eq:E-nj-ours}),
and $\hbar_{\chi}\propto\chi^{-1/2}$, the timescale of the evolution
of a quantum state is thus $\tau\propto\hat{H}/\hbar_{\chi}\propto\chi^{3/2}$.
\vskip4pt

Moreover, this time scaling law can also be understood from the perspective
of variable $c$. The timescale of a physical process in the open
set of a constant $\chi$ is deducible from its lengthscale $l$ (with
$l\propto\chi^{-1}$) and the effective speed of light $c_{\chi}$.
That is to say
\begin{equation}
\tau:=\frac{l}{c_{\chi}}\propto\chi^{-3/2}\label{eq:tau-scaling}
\end{equation}
in perfect agreement with Eq. \eqref{eq:tau-photon}.\vskip4pt

Finally, combining Eq. \eqref{eq:tau-scaling} with $l\propto\chi^{-1}$,
we arrive at an \emph{anisotropic} relationship between the timescale
$\tau$ and lengthscale $l$ of a given physical process
\begin{equation}
\tau\propto l^{\,3/2}\label{eq:anisotropic}
\end{equation}
This result is one of the key findings of our Letter. \footnote{There are two examples in classical physics that carry a $3/2$--exponent
hallmark: (1) Kepler's Third law, which states that the square of
a planet's orbital period $T$ is proportional to the cube of the
semi-major axis $a$ of its orbit, viz. $T^{2}\propto a^{3}$, or
$T\propto a^{3/2}$.\linebreak (2) The cosmic factor $a$ of a spatially
flat Einstein--de Sitter universe evolves as $a\propto t^{2/3}$,
or $t\propto a^{3/2}$. These behaviors resemble the anisotropic time
scaling, $\tau\propto l^{3/2}$, discussed in this Letter.}\vskip8pt

\textbf{\emph{Prediction: A new time dilation effect (of the Third
kind}})---The time scaling law \eqref{eq:anisotropic} indicates
that the rate of any clock---be it mechanical, electronic, or atomic---varies
in spacetime, as a function of $\chi$, in an \emph{anisotropic} fashion
in comparison to the length of a rod. Mathematically, while the length
of a rod scales as $l\propto\chi^{-1}$, the rate of a clock scales
as $\tau^{-1}\propto\chi^{3/2}$, with the $3/2$--exponent arising
due to $c_{\chi}\propto\chi^{1/2}$ (or, equivalently, due to $\hbar_{\chi}\propto\chi^{-1/2}$).
\vskip4pt

In principle, this effect can be measured \emph{experimentally}: $\,$Prepare
two identical clocks at a location $A$. Keep one clock at location
$A$ and send the other clock to a location $B$. Suppose that the
dilaton field has different values $\chi_{A}$ and $\chi_{B}$ at
the two locations $A$ and $B$, respectively. At their respective
locations, the clocks would run at different rates given by
\begin{equation}
\tau_{A}^{-1}\propto\chi_{A}^{3/2};\ \ \ \tau_{B}^{-1}\propto\chi_{B}^{3/2}\label{eq:two-clocks}
\end{equation}
When the clock from location $B$ is brought back to location $A$,
it will have registered \emph{a different amount of elapsed time}
compared to the clock that resides at location $A$ throughout the
experiment, thereby resulting in a new time dilation effect.\vskip4pt

This \emph{predicted} effect differs from the time dilation effect
in GR which is associated with the $g_{00}$ component of the spacetime
metric \citep{HafeleKeating-1972-a,HafeleKeating-1972-b}. It arises
from the dependence of the clock rate on the dilaton field $\chi$,
per Eq. \eqref{eq:two-clocks}. \footnote{In addition to the time dilation effect in GR, there is a well-known
effect in Special Relativity where two twice-intersecting time-like
paths can have different total amounts of proper time in between.
Therefore, we refer to our predicted time dilation effect as ``the
Third kind''. We emphasize that this new effect is \emph{physical}, meaning
it is, in principle, measurable \emph{by comparing the time lapses of two
clocks situated at two separate locations with different values of
the dilaton field}. We investigate this distinct phenomenon
further elsewhere.}\vskip8pt

\textbf{\emph{Revisiting the Fujii--Wetterich scheme}}---We must
emphasize that the dependence of clock rates on $\chi$ has been \emph{documented}
in the works of Fujii and Wetterich \citep{Fujii-1982,Wetterich-1988a,Wetterich-1988b,Wetterich-2013a,Wetterich-2013b,Wetterich-2014},
although it was not a focal point in their analysis. However, their
findings significantly diverge from ours. Specifically, they reported
the following relation for the clock rate \footnote{We find it illuminating to quote Fujii \citep{Fujii-1982}: \emph{``...
the time and length in the microscopic unit frame are measured in
units of $m^{-1}(t)$, in agreement with the physical situation that
the time scale of atomic clocks, for example, is provided by the atomic
levels which are determined by the Rydberg constant $(me^{4})^{-1}$''}
and Wetterich \citep{Wetterich-2013a}: \emph{``The clock provided
by the Hubble expansion in the standard description is now replaced
by a clock associated to the increasing value of $\chi$''}.}
\begin{equation}
\tau_{\,\text{FW}}^{-1}\propto\chi\label{eq:FW-tau-prediction}
\end{equation}
This contrasts with our result $\tau^{-1}\propto\chi^{3/2}$, per
Eq. \eqref{eq:tau-scaling}.\vskip4pt

Their result can be deduced as follows. If our analysis is repeated
using the FW scheme (which mandates that $\hbar$ and $c$ be constant),
utilizing Eq. \eqref{eq:e-m-scheme1}, one would obtain for the Bohr
radius
\begin{equation}
a_{B}=\frac{\hbar}{\alpha m_{\chi}c}=\frac{1}{\alpha\mu\chi}
\end{equation}
and for the energy level
\begin{equation}
E_{n}^{j}=\mathcal{N}_{n}^{j}\,m_{\chi}c^{2}=\mathcal{N}_{n}^{j}\,\mu\hbar c\,\chi\,.
\end{equation}
These expressions are identical to our results as given in Eqs. \eqref{eq:Bohr-radius-ours}
and \eqref{eq:E-nj-ours}, recalling that $\hbar$ and $c$ are constants
for the FW scheme. However, the frequency of the photon emitted during
the transition of a hydrogen atom from the excited state $\left|n=2,\,j=3/2\right\rangle $
to the groundstate $\left|n=1,\,j=1/2\right\rangle $ is
\begin{equation}
\nu_{\,\text{FW}}=\frac{\Delta E}{\hbar}=\Bigl(\mathcal{N}_{n=2}^{j=3/2}-\mathcal{N}_{n=1}^{j=1/2}\Bigr)\,\mu c\,\chi
\end{equation}
which results in $\tau_{\,\text{FW}}:=\nu_{\,\text{FW}}^{-1}\propto\chi^{-1}$,
the result stated in Eq. \eqref{eq:FW-tau-prediction}.\vskip4pt

Therefore, despite starting from the same matter Lagrangian $\mathcal{L}_{\text{QED}}^{(\chi)}$,
our scheme and the FW scheme are \emph{not physically equivalent}.
They produce \emph{two decisively different predictions} regarding
the behavior of clock rates. Future technologies may be able to distinguish
the two predictions and determine the validity of each scheme.\vskip8pt

\textbf{\emph{Advantages of our scheme}}---There are three distinct
benefits that our scheme offers as compared to the FW scheme:\vskip4pt

1. \emph{Equal treatment of mass and charge:} The FW scheme treats
particle masses and charges at \emph{disparity}: while masses are
promoted to scalar fields proportional to $\chi$, charges are treated
as parameters (per Eq. \eqref{eq:e-m-scheme1}). In contrast, our
scheme considers both inertial mass and gauge charge---intrinsic
properties of a particle---on equal footing as parameters rather
than fields.\vskip4pt

2.\emph{ Equal treatment of $G$ and $e$:} In the FW scheme, $G\propto\chi^{-2}$;
however, in our scheme, $G$ is a parameter on equal footing with
the gauge charge $e$. Consequently, our scheme suggests a commonality
between gravitational and gauge interactions.\vskip4pt

3. \emph{Universal impact on all particle types:} The FW scheme applies
only to massive particles, leaving massless particles unaffected.
In contrast, our scheme allows the dilaton---through its role in
determining $\hbar$ and $c$---to influence both massive and massless
particles equally.

\vspace{0.25cm}

\textbf{\emph{Discussions}}---Scalar degrees of freedom naturally
emerge in various theoretical frameworks, including Kaluza--Klein,
string theory, and braneworld scenarios \citep{Fujii-book}. By allowing
matter to couple non-minimally with gravity via a scalar field (e.g.,
$\chi$), scalar--tensor theories can acquire scale symmetry, enabling
them to evade observational constraints related to the fifth force
\citep{Blas-2011,Ferreira-2017}. On the ground of dimensionality,
it can be deduced that the lengthscale $l$ of a given physical process
is determined by the value of $\chi$, per $l\propto\chi^{-1}$. This
relationship justifies the use of the term ``dilaton'' for the field
$\chi$. \vskip4pt

In this Letter, we imposed two requirements: (i) A dilaton field $\chi$
exists and directly couples with matter; and (ii) The intrinsic properties
of matter---i.e., its gauge charge and inertial mass---are independent
of the dilaton. Using the Lagrangian $\mathcal{L}_{\text{QED}}^{(\chi)}$
as a prototype, we demonstrated that these two criteria unambiguously
lead to the following dependencies of $\hbar$ and $c$ on $\chi$:
$\hbar_{\chi}\propto\chi^{-1/2}$ and $c_{\chi}\propto\chi^{1/2}$.\vskip4pt

Based on these relationships, we then established that the timescale
$\tau$ of a given physical process is related to the value of $\chi$
as $\tau\propto\chi^{-3/2}$. Together with $l\propto\chi^{-1}$,
this leads to a universal anisotropic scaling law $\tau\propto l^{3/2}$
between the timescale and lengthscale of a physical process, resulting
in a new time dilation effect. (Note: notwithstanding the resemblance
of our anisotropic timescaling law to that in Ho\v{r}ava--Lifshitz
gravity \citep{Horava-2009}, Lorentz symmetry is not broken in our
approach which respects both general covariance and local Lorentz
invariance.) \vskip4pt

Three remarks are in order. Firstly, although $h_{\chi}$ and $c_{\chi}$
vary alongside $\chi$ in spacetime, the product $\hbar_{\chi}c_{\chi}$
remains constant. Secondly, in place of the QED prototype considered
in this Letter, an extension that incorporates the Higgs field and
spontaneous symmetry breaking was presented in Ref. \citep{Nguyen-VSL1}.
Its generalization to cover the Glashow--Weinberg--Salam model operating
under electroweak symmetry is also straightforward.\vskip4pt

Finally, and most importantly, the stated dependencies of $\hbar$
and $c$ on $\chi$ are derived \emph{solely from the matter Lagrangian}
without requiring detailed knowledge of the dynamics of $\chi$. This
feature renders our derivation particularly powerful and versatile:
$\,$\emph{the key requirement is the existence of a dilaton that
directly couples with matter}, whereas the gravitational Lagrangian
$\mathcal{L_{\chi}}$, which governs the dynamics of the dilaton,
does not play any essential role in our derivation. This versatility
opens the door to exploring other candidates for the gravitational
Lagrangian in future research, all while maintaining the ubiquity
of variable $\hbar$ and $c$.\vskip8pt

\textbf{\emph{Physical consequences}}---A varying $c$ in spacetime
leads to direct and significant ramifications in theoretical and observational
cosmology. Specifically, a variation in $c$ affects the propagation
of light rays in an expanding universe, thereby altering standard
cosmography both qualitatively and quantitatively. In \citep{Nguyen-VSL2},
we consider a VSL cosmology that relates the speed of light to the
cosmic scale factor as $c\propto a^{-1/2}$. For this cosmology, we
find that the canonical Lema\^itre redshift formula, $1+z=a^{-1}$,
is no longer applicable and should be modified to $1+z=a^{-3/2}$.
\emph{The $3/2$--exponent in this new formula arises from the anisotropic
time scaling discussed in this Letter.} Consequently, our new Lema\^itre
redshift formula necessitates a reanalysis for the Hubble diagram
of Type Ia supernovae in the context of late-time accelerating expansion.
In \citep{Nguyen-VSL2}, we conduct this reanalysis for the Pantheon
Catalog using our VSL cosmology instead of the concordance $\Lambda$CDM
model, and provide a new robust explanation for late-time cosmic acceleration
based on VSL, while bypassing the need for dark energy.\vskip6pt

It is important to note that the observational bounds established
in the literature in support of a constant speed of light have predominantly
relied on standard cosmology \citep{Abdo-2009,Agrawal-2021,Cai-2016,Cao-2017,Colaco-2022,Qi-2014,Ravanpak-2017,Salzano-2016a,Salzano-2016b,Santos-2024,Uzan-2003,Uzan-2011,Wang-2019,Zhang-2014,Cao-2018,Zou-2018,Mendonca-2021}.
However, our new Lema\^itre redshift formula represents a critical
departure from this framework, thereby warranting a reevaluation of
these constraints. The consensus regarding the absence of variation
in $c$ in observational cosmology must be reconsidered in light of
our findings, prompting a comprehensive reanalysis of these established
bounds.\vskip6pt

Additionally, a varying $\hbar$ may have significant implications
for quantum fields in curved spacetimes. One promising avenue for
future research on this front is the thermodynamics of black holes
in scale-invariant gravity.\vskip8pt

\textbf{\emph{Conclusion}}---Symmetries play an instrumental role
in constructing modern physical theories. Prominent
examples are general covariance in GR and gauge invariance in the
Glashow--Weinberg--Salam model of particle physics. In this Letter,
we examined a \emph{scale-invariant} action specified by Eqs. \eqref{eq:L-chi},
\eqref{eq:L-chi-QED} and \eqref{eq:S-2}, which is generally covariant,
\emph{locally} Lorentz invariant, and $U(1)$ gauge invariant (for
the matter sector). The scale invariance necessitates a non-minimal
coupling of matter to gravity via a dilaton. Although this theory has been extensively investigated \citep{Ferreira-2017,Fujii-1982,Wetterich-1988a,Wetterich-1988b,Wetterich-2013a,Wetterich-2013b,Wetterich-2014,Blas-2011,Ghilencea-2019,Kannike-2017,Karananas-2016,Nishino-2011,Rubio-2014,Rubio-2017,Salvio-2014,Shaposhinikov-2011,Einhorn-2017,Bardeen-1995},
its capacity to permit variable Planck constant and speed of light in \emph{curved}
spacetimes has been overlooked in the literature. We addressed this
shortcoming by providing---for the first time---a concrete mechanism
to generate variations in $\hbar$ and $c$ across spacetimes, leading
to material consequences in both theoretical and observational cosmology,
as presented in Ref. \citep{Nguyen-VSL2}.
\begin{acknowledgments}
I thank Clifford Burgess, Tiberiu Harko, Robert Mann and Anne-Christine
Davis for their constructive and supportive comments during the development
of this work. I thank the anonymous Referee for his or her significant
feedback in improving this Letter.
\end{acknowledgments}

\end{document}